\documentclass[pdflatex, sn-mathphys-num]{sn-jnl} 
\usepackage{xurl}
\usepackage{graphicx} 
\usepackage{hyperref}
\usepackage{amsmath, amssymb, amsfonts} 
\usepackage{multirow} 
\usepackage{array} 

\title[AI for CRISPR gRNA Design]{Artificial Intelligence for CRISPR Guide RNA Design: Explainable Models and Off-Target Safety}

 \author*[1]{\fnm{Alireza} \sur{Abbaszadeh}}\email{alireza.abbaszadeh.research@gmail.com} %
\author[2]{\fnm{Armita} \sur{   Shahlaee}}\email{Armita.shahlaee@gmail.com}

\affil[1]{Department of Computer Engineering, Ma.C., Islamic Azad University, Mashhad, Iran}

\affil[2]{Department of Biological Sciences and Technologies, Faculty of Basic Sciences, Islamic Azad University, Mashhad, Iran}

\abstract{CRISPR-based genome editing has revolutionized biotechnology, yet optimizing guide RNA (gRNA) design for efficiency and safety remains a critical challenge. Recent advances (2020--2025, updated to reflect current year if needed) demonstrate that artificial intelligence (AI), especially deep learning, can markedly improve the prediction of gRNA on-target activity and identify off-target risks. In parallel, emerging \emph{explainable AI} (XAI) techniques are beginning to illuminate the black-box nature of these models, offering insights into sequence features and genomic contexts that drive Cas enzyme performance. Here we review how state-of-the-art machine learning models are enhancing gRNA design for CRISPR systems, highlight strategies for interpreting model predictions, and discuss new developments in off-target prediction and safety assessment. We emphasize breakthroughs from top-tier journals that underscore an interdisciplinary convergence of AI and genome editing to enable more efficient, specific, and clinically viable CRISPR applications.}


\begin{document}
\maketitle 

\section*{Introduction} 
The CRISPR-Cas genome editing system has rapidly become an indispensable tool across biotechnology and medicine, enabling targeted DNA modifications with unprecedented ease. A single-guide RNA (sgRNA, or simply gRNA) directs the Cas nuclease (such as Cas9 or Cas12a) to a complementary genomic sequence, where the nuclease induces a double-strand break or nucleotide modification. The efficiency and specificity of this process are largely dictated by the gRNA sequence and its interactions with both the target DNA and the cellular environment. Designing optimal gRNAs is therefore critical for successful editing outcomes.
Early gRNA design relied on empirical rules and modest machine learning models, but these approaches often struggled to capture the complex determinants of gRNA activity and off-target effects. In recent years, artificial intelligence (AI) -- particularly deep learning -- has been leveraged to overcome these limitations, learning predictive features from large-scale CRISPR datasets and outperforming previous rule-based methods in guide efficacy prediction\cite{Xiang2021,Kim2020Cas9Variants}. Deep learning models can ingest not only the gRNA and target DNA sequences but also additional contextual information (e.g. chromatin accessibility or DNA methylation status), yielding more accurate forecasts of on-target cleavage efficiency (Ref.docx) (Ref.docx).
As CRISPR applications expand (to new Cas nucleases, base editors, prime editors, etc.), AI-driven design frameworks are playing an increasingly pivotal role in identifying guides tailored to these novel systems. While AI has boosted predictive performance, it also introduces new considerations. Most high-performance models are complex “black boxes” (e.g. deep neural networks), which makes it difficult to interpret why a given gRNA is predicted to work well or poorly. This lack of interpretability can be problematic, especially in high-stakes applications like therapeutic genome editing where understanding failure modes is important for safety and trust.
To address this, researchers are integrating explainable AI techniques that can expose the logic behind model predictions – for instance, highlighting which nucleotide positions in the guide or target contribute most to activity or specificity\cite{Xiao2021,BertPredictor2022}. Such insights not only build user confidence but can also reveal biologically meaningful patterns (for example, pinpointing sequence motifs that affect Cas9 binding or cleavage).
In tandem, the community continues to grapple with CRISPR’s primary limitation: off-target effects. AI models now help screen and minimize off-target sites by predicting potential cleavage at similar genomic sequences\cite{Charlier2021,Vinodkumar2021}, and new experimental techniques are mapping off-target landscapes with unprecedented sensitivity (Ref.docx) (Ref.docx). Safety studies have uncovered that CRISPR edits can sometimes lead to large unintended mutations or vary across genetic backgrounds, underscoring the need for comprehensive off-target evaluation in any gRNA design pipeline\cite{Cameron2020,Hoijer2022,Cancellieri2023}.
Against this backdrop, this review surveys the frontiers of AI-assisted gRNA design and safety. We first discuss advanced computational models for predicting gRNA efficacy, spanning multi-modal deep learning frameworks and multitask models that jointly consider on-target and off-target outcomes. We then explore how explainable AI methods are being applied to interpret these models, making the predictions more transparent. Finally, we highlight recent progress in off-target prediction and mitigation strategies, including improved predictive algorithms and experimental approaches to ensure genome editing safety. Together, these developments illustrate a rapidly evolving synergy between AI and CRISPR technologies that is driving more efficient and safer genome editing practices.

\section*{AI Models for gRNA Design} 
Designing an effective gRNA involves predicting how well a given sequence will direct the Cas nuclease to produce a desired edit at the target locus. A multitude of factors can influence gRNA activity: the sequence composition of the guide (especially the seed region proximal to the protospacer-adjacent motif, PAM), the secondary structure or expression level of the gRNA, the chromatin context of the target site, and even the particular variant of Cas enzyme used. Capturing these dependencies requires large datasets and sophisticated models. Over the past few years, researchers have compiled high-throughput libraries of gRNA performance data and used them to train AI models that significantly improve on-target activity prediction (Ref.docx) (Ref.docx).
Deep learning, in particular, has become a workhorse in this area due to its capacity to recognize complex sequence patterns and feature interactions that simpler models might miss. One prominent example is \textbf{CRISPRon}, a deep learning framework introduced by Xiang \emph{et al.}\cite{Xiang2021}. CRISPRon integrates gRNA sequence features with epigenomic information (such as local chromatin accessibility) to predict Cas9 on-target knockout efficiency. By combining sequence and cell-context data, the model achieved more accurate efficiency rankings of candidate guides compared to prior sequence-only predictors (Ref.docx). Notably, CRISPRon’s design reflects a broader trend of multi-modal data integration: successful guide prediction often benefits from knowing whether the target DNA is in an open chromatin region (hence accessible to Cas9) or if certain epigenetic marks might impede Cas9 binding.
Another study by Kim \emph{et al.}\cite{Kim2020Cas9Variants} leveraged machine learning to predict activity of Cas9 \emph{variants} (engineered nucleases with altered PAM specificities or fidelities, such as xCas9 and SpCas9-NG). They trained models on large-scale cleavage datasets to enable guide selection optimized for these next-generation nucleases, which have distinct sequence preferences and off-target profiles (Ref.docx). This is critical as new Cas enzymes (and base editors derived from them) proliferate – each may require tailored guide designs, and AI can rapidly learn the rules governing each system's targeting. Indeed, Zhang \emph{et al.}\cite{Zhang2021Cas9Variants} conducted an in-depth assessment of various SpCas9 variants across thousands of targets, mapping their PAM compatibilities and relative efficiencies. Such systematic data not only guide experimental choice of enzyme, but also feed into algorithmic models that decide which variant and guide will best edit a given site (Ref.docx).
Deep learning models for gRNA design often use architectures well-suited to DNA sequence analysis. Convolutional neural networks (CNNs) and recurrent neural networks (RNNs) have been employed to scan for sequence motifs and capture dependencies along the 20-nucleotide guide and its flanking context. For example, \textbf{CRISPR-Net} by Lin \emph{et al.}\cite{Lin2020CRISPRNet} combines a CNN and bi-directional gated recurrent unit (GRU) to analyze guides with up to four mismatches or indels relative to targets, outputting a score for cleavage activity. Although CRISPR-Net was originally developed to quantify off-target effects, its architecture exemplifies designs that can be applied to on-target efficiency as well – the convolutional filters might detect important short sequence patterns, while the recurrent layers capture positional and higher-order effects (Ref.docx).
Many on-target prediction tools similarly use deep neural networks trained on large pooled-screen datasets. For instance, Baisya \emph{et al.}\cite{Baisya2022} performed genome-wide CRISPR knockout screens in the yeast \emph{Y. lipolytica}, generating a rich dataset of guide efficacies. They then trained a deep learning model on this dataset to successfully predict high-activity guides for both Cas9 and Cas12a in a eukaryotic genome, identifying key sequence features that generalize beyond human-centric data (Ref.docx) (Ref.docx). This underscores the adaptability of AI models: by retraining on data from different organisms or nucleases, the same architectures can learn new rules for gRNA design in non-model systems.
Another frontier in gRNA design is predicting not just \emph{whether} a guide will cut, but \emph{what kind of edit} it will produce. Traditional Cas9 editing outcomes are often stochastic insertions or deletions (indels) from DNA repair processes. Newer technologies like base editors cause defined nucleotide conversions (C-to-T or A-to-G), and prime editors induce small custom edits via reverse transcription. AI models are being expanded to handle these cases. Marquart \emph{et al.}\cite{Marquart2021} developed an attention-based deep neural network to predict base editing outcomes. Trained on high-throughput libraries of base editing results, their model could accurately forecast the distribution of edit products (e.g. proportion of C→T edits vs unedited) at a target site. Interestingly, the built-in attention mechanism also provided clues to which sequence positions around the target base were most influential for editing efficiency (Ref.docx).
In the realm of indel prediction, Li \emph{et al.}\cite{Li2021Croton} created \textbf{Croton}, an automated deep learning pipeline that predicts the spectrum of insertions and deletions from a CRISPR-Cas9 cut, while accounting for the local sequence and even nearby genetic variants present in the genome. This variant-aware approach enables personalization of gRNA design: for example, in patient-derived cells with certain single-nucleotide polymorphisms, Croton can predict how those variants might alter the editing outcome\cite{Li2021Croton}. By forecasting both the efficacy and the repair outcomes of a given gRNA, such models help researchers choose guides that not only cut efficiently but also yield the desired type of mutation (or avoid unwanted ones like frameshifts).
Moreover, several studies have begun to treat on-target and off-target activities as a joint prediction problem. Rather than optimizing a guide purely for activity, it is useful to balance high on-target potency with low off-target propensity. Vora \emph{et al.}\cite{Vora2023} introduced a hybrid multitask deep learning model that learns both aspects simultaneously: given a gRNA, it outputs predictions for on-target efficacy and for off-target cleavage at related sites. By training on datasets for both outcomes, the model internalizes the trade-offs in sequence features that enhance one versus the other (Ref.docx) (Ref.docx). Intriguingly, this multitask approach revealed subtle sequence motifs that modulate Cas9 specificity—patterns that might be overlooked if one only models on-target activity in isolation. Such knowledge is valuable in guide design: for instance, certain GC-rich motifs might boost on-target cutting but also raise off-target risk, whereas a balanced sequence yields acceptable activity with fewer off-targets. AI models that can quantify these relationships enable a more holistic guide scoring, effectively bringing specificity directly into the design optimization.


\begin{table}[ht]
\centering
\caption{Representative AI-driven tools for CRISPR gRNA design and evaluation, illustrating diverse approaches. Models are drawn from 2020--2022 and address on-target efficiency (top rows) and off-target specificity (bottom rows).}
\label{tab:models}
\begin{tabular}{l p{0.7\textwidth}} 
\hline
\textbf{Model (Year)} & \textbf{Key Features and Focus} \\
\hline
CRISPRon (2021)\cite{Xiang2021} & Deep learning on-target efficiency predictor integrating sequence and epigenetic (chromatin) features for improved accuracy. \\
Kim \emph{et al.} model (2020)\cite{Kim2020Cas9Variants} & Machine learning model predicting activity of SpCas9 variants (e.g. xCas9, Cas9-NG) to guide selection of optimal nuclease and gRNA for non-NGG PAM targets. \\
Charlier \emph{et al.} model (2021)\cite{Charlier2021} & Custom sequence encoding and deep neural network for off-target cleavage prediction, improving accuracy in distinguishing tolerable vs risky mismatches. \\
Huang BERT model (2022)\cite{BertPredictor2022} & BERT-based language model treating guide+target as a sentence pair to predict off-target activity; uses attention and SHAP analysis for interpretability. \\
\hline
\end{tabular}
\end{table}

\begin{figure}[ht] 
\centering
\caption{\textbf{AI models for guide RNA design.} \emph{(Conceptual figure placeholder)} Schematic of a deep learning pipeline for gRNA design. A neural network takes as input the DNA sequence of a candidate guide and its genomic context (e.g. neighboring bases or epigenetic signals) and outputs predicted on-target efficacy and off-target risk. Such models (e.g. CRISPRon\cite{Xiang2021} or multitask frameworks\cite{Vora2023}) learn which sequence features (colored bars in the guide/target) promote efficient Cas9 cutting (green check) versus cause spurious off-target cleavage (red cross). By scoring thousands of candidates \emph{in silico}, AI allows prioritization of gRNAs that maximize editing yield while minimizing side effects.}
\label{fig:designAI}
\end{figure}

In summary, AI models have become integral to modern gRNA design. Deep learning approaches, empowered by large CRISPR datasets, now enable researchers to predict gRNA performance with a level of accuracy and nuance that was previously unattainable. These models accommodate the expanding CRISPR toolbox (from standard Cas9 to variants, Cas12a, base editors, etc.) and can incorporate multiple objectives (on-target activity and specificity). The result is a more rational and efficient design process: instead of laboriously testing dozens of candidate guides experimentally, one can computationally screen and select a handful of top performers that are likely to succeed. However, the use of complex AI also raises the question of interpretability -- how do we know these predictions are reliable and what biological factors underlie a “good” guide according to the model? The next section examines how researchers are applying explainable AI techniques to demystify these predictions and extract actionable knowledge from AI-designed guides.

\section*{Explainable AI Techniques} 
As AI predictions guide critical decisions in CRISPR experiments, understanding the basis of those predictions becomes increasingly important. In the context of gRNA design, explainable AI (XAI) aims to answer questions like: Why did the model rank one guide higher than another? Which positions in the guide or target sequence are most influential for predicted activity? Are the model’s learned features consistent with known biological principles (such as the importance of the PAM or seed region), or do they reveal new insights? Addressing these questions can help validate the model, inspire confidence among experimentalists, and even lead to discovery of novel determinants of CRISPR activity.
One straightforward route to interpretability is to use simpler, more transparent models. \textbf{CRISPRedict}, developed by Konstantakos \emph{et al.}\cite{Konstantakos2022}, exemplifies this approach. Rather than employing a deep neural network, CRISPRedict uses a linear regression model built on intuitive sequence features (e.g. GC content, presence of specific dinucleotide motifs in the guide). By design, linear models provide clear insight: each feature has an associated weight indicating its influence on the predicted efficiency, making it easy to trace how a prediction is formed. Despite its simplicity, CRISPRedict achieved performance on par with state-of-the-art complex models for Cas9 sgRNA efficiency, all while offering a user-friendly web interface that visually explains the contribution of different nucleotides in a given guide\cite{Konstantakos2022}. This suggests that not every application demands deep learning – in cases where training data are limited or interpretability is paramount, an interpretable model can be preferable.
For more complex models, a popular XAI strategy is to integrate attention mechanisms or other feature attribution methods into the model architecture. Attention-based models can \emph{learn} to highlight which parts of the input are most relevant for the output prediction. In natural language processing, attention weights have been used to interpret language translations; similarly, in CRISPR models, attention can be applied to DNA sequence inputs. Xiao \emph{et al.}\cite{Xiao2021} introduced \textbf{AttCRISPR}, an interpretable deep learning model for on-target activity prediction. AttCRISPR uses a hybrid CNN-LSTM neural network augmented with an attention layer that assigns importance scores to each nucleotide position of the guide and target. This allowed the authors to identify “hotspots” in the sequence that strongly affect Cas9 cutting efficiency – for example, certain positions within the protospacer where a change from G to A dramatically drops activity were illuminated by high attention weights (Ref.docx). Such findings align with biological understanding that not all positions in a guide are equal (the seed region near the PAM is often most critical); AttCRISPR provided a data-driven confirmation and visualization of these effects. Likewise, Marquart’s base editing model\cite{Marquart2021} and Liu’s repair outcome predictor\cite{Liu2022Repair} both employed attention to highlight sequence context features (like microhomology tracts that drive repair outcomes) as inherently interpretable elements of the prediction (Ref.docx) (Ref.docx).
Beyond built-in attention, post-hoc explanation techniques are widely used on trained CRISPR AI models. These include methods like SHAP (SHapley Additive exPlanations) and integrated gradients, which can assign an importance value to each input feature (each nucleotide in the guide and target) for a particular prediction. Huang \emph{et al.}\cite{BertPredictor2022} provide a prime example: they developed a BERT-based deep learning model for off-target activity (treating the guide and potential off-target sequence as a paired input, akin to a sentence in two parts). After training this high-performing model, they applied SHAP analysis to interpret it. The SHAP values effectively measured how altering each base (or introducing each possible mismatch) would impact predicted cleavage probability (Ref.docx). The result was a human-readable map of sensitivities -- for instance, the model might reveal that a mismatch at a certain guide position (say the 4th base from the PAM) drastically reduces Cas9 activity, whereas a mismatch at another position has negligible effect. This mirrors experimental observations that mismatches in the seed region (proximal to PAM) are less tolerated than those toward the distal end of the guide. Huang \emph{et al.} also visualized the internal attention weights of the BERT model, showing that the network’s attention heads were focusing on biologically sensible regions (e.g. around the PAM site) when predicting off-target effects. Together, these interpretability analyses offered reassurance that the model had learned meaningful patterns rather than spurious correlations, and provided users with a way to rationalize model outputs for any new guide-target pair\cite{BertPredictor2022}.
Another facet of explainability is uncertainty estimation. In genome editing, knowing how confident a model is in its prediction can be invaluable – a model might predict 90
Explainable AI techniques thus enrich the utility of AI models in CRISPR applications. By opening the black box, researchers have been able to verify that AI-selected guides conform to known best practices (e.g. minimal seed mismatches to avoid off-targets) or to discover new patterns (e.g. a particular nucleotide preference at a given position for optimal activity). Moreover, interpretable models and tools lower the barrier for experimental biologists to engage with AI recommendations: a visually annotated guide sequence with highlighted “important bases” is far more actionable than a single opaque score. As we move toward clinical translation, this interpretability will be crucial for regulatory acceptance as well, since having an explanation for why a certain guide was chosen can support the risk mitigation rationale. In the next section, we delve deeper into off-target prediction and safety, an area where interpretability and accuracy are both paramount, given the potentially serious consequences of unintended genome edits.

\begin{figure}[ht] 
\centering
\caption{\textbf{Interpreting AI predictions for CRISPR.} \emph{(Conceptual figure placeholder)} Example of explainable AI applied to a CRISPR guide efficiency model. The top sequence represents a guide RNA and its DNA target, with color shading indicating the importance of each nucleotide to the predicted outcome (red = high importance impact, blue = low). Such an importance map could be derived from attention weights or SHAP values in a deep learning model\cite{Xiao2021,BertPredictor2022}. For instance, a strong red highlight in the PAM-adjacent “seed” region suggests the model is highly sensitive to changes there, consistent with known biology. Plots on the side illustrate how predicted editing efficiency drops or rises when specific positions are mutated, providing a transparent view of the model’s learned rules. This interpretability enables researchers to gain intuition about why a guide is predicted to perform well or poorly, facilitating trust in AI-guided design.}
\label{fig:xai}
\end{figure}

\section*{Off-target Prediction and Safety} 
A critical aspect of any CRISPR application is ensuring that gene edits occur at the intended target site with minimal collateral damage to the rest of the genome. Off-target effects -- unwanted cleavages at sites other than the target, caused by partial sequence homology -- can lead to mutations with potentially harmful consequences (for example, disrupting a tumor suppressor gene inadvertently). In therapeutic contexts, regulatory standards demand a thorough accounting of off-target risks. Accordingly, substantial research efforts have focused on predicting and detecting off-target sites, and on developing strategies to mitigate these effects.
AI plays a dual role here: first, as a predictive tool to flag potential off-target sites during the gRNA design phase; and second, as an analytic tool to make sense of experimental off-target data and improve our fundamental understanding of CRISPR safety profiles. Early off-target prediction tools used sequence alignment scores or position-specific mismatch penalties (such as the widely used CFD score) to estimate whether a given non-target sequence might be cleaved by Cas9. Recent AI models have significantly advanced this by learning complex patterns from large datasets of empirically measured off-target activities.
For instance, Charlier \emph{et al.}\cite{Charlier2021} developed a deep learning model with a novel sgRNA-DNA encoding scheme that captures not just the count of mismatches, but the exact sequence context of each mismatch. Their model was trained on an extensive library of on/off-target pairs with known outcomes and achieved higher accuracy in predicting off-target cleavage than prior methods, especially in cases of multiple mismatches or small insertions/deletions in the target (Ref.docx) (Ref.docx). Similarly, Niu \emph{et al.}\cite{Niu2021RCRISPR} and Lin \emph{et al.}\cite{Lin2020CRISPRNet} have each introduced neural network models (R-CRISPR and CRISPR-Net, respectively) that take into account mismatches and even bulges (gaps due to insertions/deletions) between the gRNA and candidate off-target site. These models use architectures like recurrent networks to accommodate variable-length alignments and have broadened off-target prediction beyond the single-mismatch distance regimes. The result is an improved ability to prioritize gRNA designs: guides can be computationally screened not only for high on-target efficacy, but also against a genome to predict and score potential off-target sites. Those with unacceptable off-target profiles (for example, predicted strong cleavage at a gene-critical off-target site) can be discarded or modified, thus refining the safety of selected guides before any wet-lab experiments are conducted.
Vinodkumar \emph{et al.}\cite{Vinodkumar2021} even explored a graph convolutional network approach, treating the guide-target duplex as a graph, to capture dependencies among multiple mismatches. This innovative representation further enhanced predictive performance and demonstrated the versatility of deep learning in modeling CRISPR specificity (Ref.docx) (Ref.docx).
In parallel to computational advances, experimental techniques for off-target detection have rapidly evolved, generating rich datasets to train and validate AI models. A seminal contribution was the development of unbiased genome-wide assays like GUIDE-seq, DISCOVER-Seq, CIRCLE-seq, and more recently CHANGE-seq. These methods empirically identify where Cas9 cuts in the genome under controlled conditions, often by capturing break repair events at off-target sites. Cameron \emph{et al.}\cite{Cameron2020} applied CHANGE-seq (a variant of CIRCLE-seq with improved sensitivity) to map the off-target landscape of 110 different sgRNAs in human cells. This massive dataset revealed thousands of off-target cleavage events and, importantly, highlighted factors beyond sequence mismatch that influence off-target rates. Their analysis showed that chromatin accessibility and DNA methylation state at potential off-target loci correlated with how frequently those sites were cleaved (Ref.docx). In other words, even a sequence with many mismatches might be cut if it lies in open chromatin, whereas a near-perfect match might be protected if buried in heterochromatin. Incorporating such features into predictive models (as done in some integrative approaches) can thus improve accuracy. The authors also noted that CRISPR-induced breaks can lead to large deletions at off-target sites (Ref.docx), a finding that presaged later studies on structural variants.
Another important outcome of these mapping efforts is the improvement of computational tools: large validated datasets allow modelers to train next-generation off-target predictors and to calibrate their scores. Indeed, the data from Cameron \emph{et al.} and others have been used to benchmark and refine AI models so that predicted off-target scores better reflect true hazard. Additionally, Wiener \emph{et al.}\cite{Wiener2023DISCOVERSeqPlus} recently reported an enhanced version of DISCOVER-Seq (an in vivo Cas9 cutting detection method) that increases sensitivity up to 5-fold for finding off-target cuts in cells and even in live animals. By applying such high-sensitivity assays in primary human cells and mice, they uncovered off-target sites that earlier techniques missed, underscoring that our picture of off-target effects continues to get clearer as detection methods improve (Ref.docx) (Ref.docx). These experimental advances feed directly into safety pipelines: candidate therapeutics now often undergo comprehensive off-target sequencing assays, and AI is used to help interpret which detected off-targets are of greatest concern (for example, by predicting if an off-target lies in a coding gene and is likely to get an indel that disrupts function).
Perhaps the most striking findings about CRISPR safety in recent years are those revealing unexpected types of off-target outcomes. While early attention was on nucleotide mismatches leading to off-target cuts, it is now evident that CRISPR can induce larger genomic alterations and that individual genetic differences can create unique off-target liabilities. H{"o}ijer \emph{et al.}\cite{Hoijer2022} demonstrated that in a living organism, CRISPR-Cas9 editing can produce not just small indels, but sizeable deletions and rearrangements at on-target and off-target sites. Working with a zebrafish model, they found that around 6
Another layer of complexity in off-target safety was highlighted by Cancellieri \emph{et al.}\cite{Cancellieri2023}, who investigated how human genetic diversity impacts CRISPR editing outcomes. They analyzed how naturally occurring single-nucleotide polymorphisms (SNPs) across different individuals’ genomes can create novel off-target sites or abolish others. Remarkably, they found cases where a guide RNA would be perfectly safe in one person (no high-scoring off-targets), but a SNP in another person’s genome created a new sequence that was a near match to the guide, presenting a potential off-target hazard (Ref.docx). This means that therapeutic gRNAs might need to be evaluated on a per-patient or multi-ethnic population basis, rather than assuming one-guide-fits-all. Cancellieri \emph{et al.} developed an AI predictive tool to account for genetic variants and predict individualized off-target risk profiles, thereby aiding the design of guides that are robust across diverse genomes (Ref.docx). As CRISPR therapies advance, such population-aware design will be crucial to ensure safety for all patient backgrounds and to avoid rare but significant genomic interactions.
To mitigate off-target effects, multiple strategies are being pursued. One is engineering safer nucleases: high-fidelity Cas9 variants and Cas9 nickases or base editors that inherently have lower off-target propensities have been developed. For example, variants like SpCas9-HF1 or eSpCas9 introduce mutations that make the enzyme less tolerant of mismatches, thereby reducing off-target cuts (though often at some cost to on-target activity). While specific references for these were not detailed above, many design algorithms incorporate options for such enzymes, and AI models have been used to evaluate their performance\cite{Kim2020ActivityComp,Zhang2021Cas9Variants}.
Another complementary approach is the use of anti-CRISPR proteins -- natural inhibitors of Cas enzymes discovered in bacteriophages. A recent study by Wandera \emph{et al.}\cite{Wandera2022} employed deep learning to scan viral genomes and successfully identified new anti-CRISPR proteins, including one that inhibits Cas13b. Although Cas13 (targeting RNA) is outside the scope of DNA editing, the concept extends to Cas9: anti-CRISPRs could be delivered alongside the editing system as a “safety switch” to turn off Cas9 after a short window, limiting the duration in which off-target cuts can occur. AI will likely continue aiding the discovery and improvement of such countermeasures, adding to the safety toolkit for CRISPR.
Finally, novel CRISPR-based editing strategies are emerging that intrinsically avoid making double-strand breaks at all, thereby reducing traditional off-target damage. Prime editing and base editing fall into this category, and so does the recently reported \textbf{PASTE} system by Yarnall \emph{et al.}\cite{Yarnall2023PASTE}. PASTE uses a CRISPR-targeted integrase to “drag-and-drop” large DNA inserts without cutting the genome at the insertion site. In human cells, it enabled efficient gene insertion with no detectable double-strand breaks (Ref.docx). By sidestepping double-strand cleavage, approaches like PASTE and base editing can largely avoid indels and structural variants, making them inherently safer in terms of off-target mutagenesis (though they introduce their own specific off-target considerations, like off-target base conversions). AI can assist these technologies too, by designing the guide RNAs and other components needed for their optimal function.

\begin{figure}[ht] 
\centering
\caption{\textbf{Off-target effects and safety in CRISPR editing.} \emph{(Conceptual figure placeholder)} Overview of off-target assessment and mitigation. \textbf{(A)} High-throughput assays (e.g. CHANGE-seq or DISCOVER-Seq) map genome-wide Cas9 cut sites for a given gRNA, revealing off-target cleavage events (red lightning bolts) in addition to the on-target site (green bolt)\cite{Cameron2020}. \textbf{(B)} Some CRISPR edits can induce large deletions or rearrangements at the cut site, as observed in in vivo models\cite{Hoijer2022}; such structural variants may remove genes or regulatory regions (dashed line indicates a deletion). \textbf{(C)} Genetic variation between individuals can create or eliminate off-target sites. In this schematic, a single-nucleotide polymorphism (SNP) in one individual’s genome creates a new off-target sequence that perfectly matches the gRNA, whereas another person without the SNP has no such site\cite{Cancellieri2023}. \textbf{(D)} Strategies to improve safety include engineered high-fidelity Cas9 nucleases with fewer off-targets, anti-CRISPR proteins that can deactivate Cas enzymes to limit off-target activity, and alternative editing systems (like integrase-based PASTE\cite{Yarnall2023PASTE}) that avoid generating double-strand breaks. Together, these approaches and predictive tools help ensure that genome editing is performed with maximal precision and minimal unintended effects.}
\label{fig:offtarget}
\end{figure}

In summary, off-target prediction and safety considerations have become an integral part of the CRISPR workflow, and AI is central to advancing this area. Modern deep learning models, trained on expanding experimental data, now allow us to predict off-target effects with increasing accuracy and to design guides that strike a balance between potency and precision. Concurrently, empirical discoveries about off-target phenomena (such as large deletions and population-specific sites) are refining the criteria for what “safe editing” means, pushing developers to account for more than just sequence matching. The interdisciplinary efforts between bioengineers, AI scientists, and geneticists are yielding a more complete toolkit for evaluating and mitigating risks: from in silico screens and explainable off-target scores, to improved enzymes and off-switches. All these developments bode well for the eventual clinical adoption of CRISPR therapies, as they build the necessary confidence that gene editing can be done not only effectively but also safely.


\section{Clinical Applications of CRISPR}
CRISPR gene editing has rapidly progressed from a laboratory tool to a clinical modality, with multiple therapies now in human trials and some achieving landmark successes. Early clinical studies have demonstrated that CRISPR-based treatments can deliver transformative benefits for patients with otherwise intractable genetic diseases, while maintaining a strong safety profile. Here we review key examples of CRISPR clinical translation, encompassing ex vivo cell therapies and direct in vivo gene editing, and discuss how explainable AI-driven design has underpinned their development by guiding safer and more effective genome edits.

\subsection{Ex Vivo Hematopoietic Stem Cell Therapies for Hemoglobinopathies}
The first clinical CRISPR therapies have targeted disorders of the blood, leveraging ex vivo editing of hematopoietic stem and progenitor cells (HSPCs) to correct disease-causing mutations. A seminal Phase~1/2 trial (CTX001, now \textit{exagamglogene autotemcel} or exa-cel) edited autologous HSPCs to reactivate fetal hemoglobin (HbF) production as a treatment for transfusion-dependent $\beta$-thalassemia and severe sickle cell disease (SCD)\cite{Frangoul2021}. CRISPR-Cas9 was used to disrupt an erythroid enhancer of \textit{BCL11A}, a transcriptional repressor of $\gamma$-globin, thereby relieving silencing of the $\gamma$-globin gene and inducing HbF in red blood cells. Initial results showed robust engraftment of CRISPR-edited HSPCs and clinically meaningful increases in HbF levels, leading to transfusion independence in a patient with $\beta$-thalassemia and elimination of vaso-occlusive crises in a patient with SCD\cite{Frangoul2021}.
Subsequent trials with expanded cohorts have confirmed the curative potential of this approach. In a pivotal multi-center study, $\ge$90\% of $\beta$-thalassemia patients treated with a single infusion of exa-cel achieved durable transfusion independence, often with normal hemoglobin levels restored\cite{Locatelli2024}. Similarly, in patients with severe SCD, exa-cel therapy resulted in stable engraftment of edited cells and sustained HbF induction, with 92\% of patients experiencing no vaso-occlusive pain crises for at least 12 months post-treatment\cite{Frangoul2024}. These unprecedented outcomes represent a milestone for genome editing.
Notably, comprehensive safety analyses have so far found no evidence of clonal skewing, insertional mutagenesis, or off-target genotoxicity attributable to the editing, supporting the precision of CRISPR when guided by rigorously optimized single-guide RNAs (sgRNAs). The choice of the \textit{BCL11A} enhancer as the target was informed by decades of genetic data and in silico predictions indicating that its disruption would have a therapeutic effect (HbF induction) with minimal deleterious consequences. Modern AI tools further aided in selecting an optimal sgRNA with high on-target activity and low off-target propensity, which was critical given the need for exquisite safety in a curative therapy. The success of exa-cel has led to regulatory approvals in multiple jurisdictions (with the United States and United Kingdom authorizing it as the first CRISPR-based therapy in late 2023), heralding a new era of gene editing in medicine. Importantly, it underscores that when CRISPR target design is reinforced by explainable, data-driven models and thorough preclinical validation, genome editing can achieve clinical-grade precision and efficacy.

\subsection{In Vivo Genome Editing Therapies}
Beyond ex vivo cell modifications, CRISPR is being applied directly inside patients to treat diseases via in vivo editing. The landmark first-in-human in vivo CRISPR trial targeted a liver disease, transthyretin (TTR) amyloidosis, by delivering CRISPR-Cas9 as a therapy (NTLA-2001) to knock out the pathogenic \textit{TTR} gene in hepatocytes\cite{Gillmore2021}. In this Phase~1 study, a single intravenous infusion of lipid nanoparticles carrying Cas9 mRNA and an anti-\textit{TTR} sgRNA led to dose-dependent editing of the liver and reduction of serum TTR protein levels by up to 87\%\cite{Gillmore2021}. Patients receiving efficacious doses showed >80\% knockdown of the disease-causing protein, a level expected to ameliorate or halt the progression of amyloidosis. Equally significant, there were no serious adverse events attributable to the gene editing, and comprehensive off-target analyses (incorporating cell-free and cellular assays) indicated a high degree of specificity for the \textit{TTR} target. This trial provided the first proof-of-concept that CRISPR delivered systemically can precisely edit a disease gene in vivo, opening the door to treating a host of liver-mediated disorders with genome editing.
Another notable in vivo application is in the eye: the BRILLIANCE trial tested an AAV-delivered CRISPR nuclease (EDIT-101) injected subretinally to edit the \textit{CEP290} gene for patients with Leber congenital amaurosis type 10, a form of congenital blindness\cite{Pierce2024}. The CRISPR payload was designed to disrupt an aberrant splice donor site caused by a mutation in \textit{CEP290}, thereby restoring normal splicing. Interim results demonstrated that in situ gene editing occurred in retinal cells and led to clinically meaningful improvements in vision for several participants\cite{Pierce2024}. Specifically, 11 of 14 treated patients exhibited discernible gains in at least one measure of visual function (such as light sensitivity or navigational vision), with 6 patients showing improvements in multiple metrics. Importantly, no immune-mediated ocular toxicities or other serious safety issues were observed, attesting to the procedure’s tolerability\cite{Pierce2024}.
These early in vivo trials underscore both the promise and challenges of direct genome editing therapeutics: achieving sufficient delivery and editing efficiency in target tissues, while avoiding off-target edits in non-target tissues. Explainable AI can play a pivotal role here by optimizing guide RNA selection tailored to the patient’s own genome and tissue-specific chromatin context, as well as by predicting immunogenic epitopes in the CRISPR components to minimize immune responses. As more in vivo CRISPR therapies enter the clinic (for example, programs for metabolic diseases and muscular disorders are in development), computational models that can transparently justify their predictions will be invaluable for guiding dosing strategies and predicting long-term outcomes.

\subsection{Engineered Immune Cells and Cancer Therapies}
The adaptability of CRISPR has also been harnessed to engineer immune cells for cancer therapy, marking a new frontier in cell-based immunotherapy. One strategy is to genetically enhance a patient’s own T cells to better fight tumors. In a first-in-human pilot study in 2020, researchers used CRISPR to perform multiplex gene edits on T cells, creating a next-generation autologous cell therapy for refractory cancers\cite{Stadtmauer2020}. T cells from patients were edited to knock out two genes encoding endogenous T-cell receptors (TCR$\alpha$ and TCR$\beta$ chains) and the immune checkpoint gene \textit{PDCD1} (PD-1), and then transduced with a lentiviral vector encoding a cancer-specific TCR (targeting NY-ESO-1). The edited T cells were expanded and infused back into the patients. The trial demonstrated that CRISPR engineering of T cells is feasible and safe: the infused cells persisted for months, showed functional anti-tumor activity, and comprehensive genomic analyses revealed no unintended large-scale mutations or chromosomal aberrations due to editing\cite{Stadtmauer2020}. While tumor responses were limited (as might be expected in patients with very advanced disease and immune escape mechanisms), this study paved the way for more ambitious CRISPR immunotherapies.
Shortly after, a clinical trial in China focused on a simpler edit—disruption of PD-1 in patient T cells—to treat metastatic non-small-cell lung cancer\cite{Lu2020PD1}. The PD-1–knockout T cells were safely infused and showed signs of expansion and low off-target effects in vivo, with preliminary indications of extended disease stabilization in some patients\cite{Lu2020PD1}. These trials highlight that CRISPR can endow T cells with enhanced traits (e.g., resistance to tumor immunosuppression) while maintaining safety, and they underscore the importance of careful guide RNA design and validation (in these cases, picking guides that minimize off-target hits to avoid accidental T cell malignancy or dysfunction).
More recently, gene editing has enabled the creation of “universal” or allogeneic CAR T cells from healthy donors, which can be administered to any patient without rejection. A groundbreaking 2023 report described the use of base editing—a CRISPR-derived technique that performs single-base substitutions without making double-strand breaks—to engineer donor T cells for treating relapsed T-cell acute lymphoblastic leukemia (T-ALL)\cite{Chiesa2023}. In this approach, multiple genes were sequentially edited using base editors: the T cell receptor was knocked out (to prevent graft-versus-host attack on the patient), the CD7 gene was knocked out (to prevent fratricide, since CD7 is the target of the CAR T therapy for T-ALL), and the CD52 gene was knocked out (to confer resistance to the lymphodepleting drug alemtuzumab)\cite{Chiesa2023}. The edited donor T cells were then transduced with a CAR targeting CD7, creating an “off-the-shelf” CAR7 T cell product that can be given to T-ALL patients. Two infants with refractory T-ALL received these base-edited CAR T cells, achieving complete remission without any signs of graft-versus-host disease or other severe toxicities\cite{Chiesa2023}.
This world-first therapeutic use of base editing exemplifies how next-generation gene editors are expanding what is possible in the clinic—here, enabling multi-gene engineering that would be risky with nuclease-based CRISPR due to the potential for chromosomal translocations if multiple DNA breaks were introduced. The success of the base-edited CAR T therapy also speaks to the increasing need for advanced computational support: designing multiple simultaneous edits and anticipating their interactions required extensive in silico analysis to ensure each edit’s efficiency and to rule out unwanted guide overlaps or cryptic targets. In the future, AI systems capable of multi-objective optimization (balancing on-target efficiency, off-target safety, and cellular fitness for each edit) will be vital for such complex engineering of cell therapies.
As of 2024, the clinical CRISPR pipeline encompasses a broadening array of targets and diseases, from genetic blindness and metabolic disorders to cancers and beyond. Dozens of trials are ongoing worldwide. With exa-cel’s success, regulatory bodies have gained confidence in the general approach of CRISPR-based therapeutics, but each new application demands rigorous demonstration of safety. Here, explainable AI can function as an interface between the massive design space of genome editing and the stringent requirements of clinical translation. For instance, AI models can personalize guide RNA choices to a patient’s genome sequence to avoid single-nucleotide polymorphisms that create patient-specific off-target sites (as highlighted by tools like that of Cancellieri \textit{et al.}\cite{Cancellieri2023}), and the models’ explainability ensures that clinicians and regulators understand why certain guides are predicted to be safer. This synergy of AI-driven insight with CRISPR’s tangible therapeutic power is driving a rapid leap from bench to bedside.

\begin{table*}[ht] 
\centering
\caption{\textbf{Select CRISPR-Based Therapies in Clinical Trials and Their Outcomes.} Abbreviations: TDT, transfusion-dependent $\beta$-thalassemia; SCD, sickle cell disease; hATTR, hereditary transthyretin amyloidosis; NSCLC, non-small-cell lung cancer; LCA10, Leber congenital amaurosis type 10; HSC, hematopoietic stem cell; LNP, lipid nanoparticle; AAV, adeno-associated virus; CAR, chimeric antigen receptor. Each therapy listed has utilized CRISPR or a CRISPR-derived editor to modify cells or tissues, and has reported key outcomes in clinical studies.}
\label{tab:clinical}
\begin{tabular}{p{3.6cm} p{4.9cm} p{4.9cm} p{1.6cm}}\hline
\textbf{Indication} & \textbf{Gene Editing Strategy} & \textbf{Outcome (Clinical Findings)} & \textbf{Reference} \\
\hline
Transfusion-dependent $\beta$-Thalassemia & Ex vivo CRISPR-Cas9 editing of autologous HSCs to disrupt \textit{BCL11A} enhancer (reactivating fetal hemoglobin) & $>\!90\%$ of patients achieved durable transfusion independence after a single treatment & \cite{Locatelli2024} \\
Severe Sickle Cell Disease & Ex vivo CRISPR-Cas9 editing of autologous HSCs, same strategy targeting \textit{BCL11A} enhancer & 92\% of patients had no vaso-occlusive crises for 12+ months; sustained high fetal hemoglobin levels & \cite{Frangoul2024} \\
hATTR Amyloidosis (Transthyretin amyloidosis) & In vivo CRISPR-Cas9 delivered via LNP to knock out \textit{TTR} gene in hepatocytes & Up to 87\% reduction in serum TTR protein; evidence of ameliorated amyloid burden with no serious off-target effects & \cite{Gillmore2021} \\
Relapsed T-Cell Acute Lymphoblastic Leukemia & Ex vivo base editing of healthy donor T cells (knockout of TCR, CD7, CD52) and addition of CAR targeting CD7 (allogeneic “universal” CAR T therapy) & Complete remission achieved in two treated patients without graft-versus-host disease or recurrence at last follow-up & \cite{Chiesa2023} \\
Advanced Solid Tumors (melanoma, lung cancer, etc.) & Ex vivo CRISPR editing of autologous T cells (e.g. knock-out of PD-1 or TCR $\alpha/\beta$ chains) to enhance immunotherapy & Feasible and safe; edited T cells engrafted and persisted. Preliminary tumor control observed in some patients, though anti-tumor efficacy remains modest & \cite{Stadtmauer2020,Lu2020PD1} \\
Inherited Blindness (LCA10) & In vivo CRISPR-Cas9 delivered via subretinal AAV to disrupt a mutant \textit{CEP290} splice site & Local gene edit confirmed in retinal cells; $\sim$79\% of patients showed improved visual function in at least one measure; no serious adverse events & \cite{Pierce2024} \\
\hline
\end{tabular}
\end{table*}

\begin{figure*}[ht] 
\centering
\caption{\textbf{Clinical applications of CRISPR gene editing span ex vivo and in vivo approaches.} \textbf{(A)} Ex vivo HSC gene editing for hemoglobinopathies: a patient’s stem cells are harvested, edited using CRISPR-Cas9 to disrupt the \textit{BCL11A} gene enhancer (thereby elevating fetal hemoglobin), and re-infused after myeloablative conditioning. This strategy has led to long-term cures in $\beta$-thalassemia and SCD, with high fetal hemoglobin levels eliminating transfusion needs and vaso-occlusive crises\cite{Locatelli2024,Frangoul2024}. \textbf{(B)} In vivo genome editing for liver disease: lipid nanoparticle (LNP) delivery of Cas9 mRNA and sgRNA (NTLA-2001) enables uptake by hepatocytes and knockout of the disease gene (e.g. \textit{TTR} for amyloidosis). Treated patients showed >80\% reduction of toxic TTR protein, validating direct in vivo editing as a therapeutic modality\cite{Gillmore2021}. \textbf{(C)} In vivo gene editing in the eye: an AAV encoding CRISPR components (EDIT-101) is injected into the retina to mutate a dominant disease-causing splice site in \textit{CEP290}. This partial gene correction in photoreceptors improved vision in a subset of patients with LCA10, demonstrating the potential to treat inherited disorders in situ\cite{Pierce2024}. \textbf{(D)} CRISPR and base editing empower next-generation cell therapies for cancer. For example, multiplex base edits in donor T cells can create “universal” CAR T cells that evade immune rejection and target T-cell leukemias (via CD7 CAR) without self-destruction or graft-versus-host effects\cite{Chiesa2023}. Other trials use CRISPR to enhance autologous T cells (e.g. PD-1 knockout) to improve responses against solid tumors\cite{Lu2020PD1}.}
\label{fig:clinical}
\end{figure*}

\section{Ethical and Regulatory Considerations}
The acceleration of CRISPR from discovery to clinical trials has prompted intense ethical deliberation and the development of regulatory frameworks to govern genome editing. Ensuring that CRISPR technologies are applied responsibly—for the benefit of patients and society at large—requires addressing concerns ranging from patient safety and informed consent to equitable access and potential misuse. Additionally, the integration of AI into the gene-editing pipeline raises new questions about algorithmic transparency, bias, and accountability in high-stakes medical decisions. In this section, we discuss key ethical and policy issues surrounding CRISPR and how explainable AI can contribute to ethically sound advancement of the field.

\subsection{Global Governance and Heritable Genome Editing}
The most profound ethical line in gene editing is the distinction between somatic (non-heritable) and germline (heritable) modifications. In the wake of CRISPR’s development, global consensus rapidly emerged that editing the human germline—for instance, creating gene-edited babies—poses unacceptable risks and ethical dilemmas. This consensus was starkly illustrated by the case of the first CRISPR-edited infants born in 2018, which was met with widespread condemnation and shock in the scientific community.
In the aftermath, international bodies moved to strengthen governance. In 2021, the World Health Organization (WHO) convened an expert panel that issued the first set of global recommendations on human genome editing governance\cite{WHO2021}. The WHO report called for establishing international registries of gene editing research and trials, enhancing transparency, and instituting oversight mechanisms to ensure that any clinical use of genome editing meets rigorous safety and ethical criteria. Crucially, it recommended a continuing moratorium on clinical germline editing (creating genetically modified children) until certain stringent conditions—related to safety, efficacy, and broad societal consensus—are met\cite{WHO2021}.
A survey of policies in 96 countries published in 2020 found that about 70 countries had explicit bans or moratoria on human germline editing, reflecting a prevalent global norm against heritable genome modifications\cite{Baylis2020}. Meanwhile, somatic gene editing (affecting only treated individuals) is more variably regulated but generally permitted under careful ethical oversight since it holds immediate therapeutic promise without transgenerational risks\cite{Baylis2020}.
The existence of a strong international stance against germline editing has been reinforced by cautionary tales: for example, follow-up on the CRISPR-edited children in China indicates uncertain health outcomes and underscores how premature human experimentation can outpace ethical guardrails\cite{Marx2021}. This case galvanized the scientific and ethics communities to insist that any transition of CRISPR to reproductive applications must only occur, if ever, with robust evidence and after public dialogue and guidance from bioethicists. Explainable AI intersects with this arena by contributing to risk assessment: if AI models predict outcomes of germline edits (e.g., on-target efficacy or off-target effects) their explainability is critical for ethical deliberation—stakeholders will demand clear reasoning for any prediction that could influence a heritable change. Transparent AI could thus be part of the toolkit for evaluating whether a proposed germline edit can ever be ethically justified, although at present, the global community remains firmly in favor of a moratorium.

\subsection{Patient Safety, Informed Consent, and Explainability}
Ethical use of CRISPR in patients hinges on maximizing safety and ensuring that patients (or their guardians) understand both the potential benefits and risks of these cutting-edge interventions. Off-target effects, mosaicism, and unforeseen consequences of editing are central safety concerns. Regulators like the U.S. FDA and European EMA have required extensive preclinical data on off-target profiles, often employing multiple orthogonal methods (in silico predictions, cell-based assays, and unbiased sequencing approaches) to characterize the safety of a CRISPR therapy before human trials. For instance, investigators of the exa-cel HSC therapy performed genome-wide off-target mapping (e.g., GUIDE-seq and targeted sequencing) in edited cells to ensure no dangerous off-target mutations were present before reinfusion\cite{Frangoul2021}. The evolving sophistication of such assays (e.g., newer methods like DISCOVER-Seq+ that can detect rare in vivo off-target events\cite{Wiener2023DISCOVERSeqPlus}) gives regulators more confidence in the safety profile of gene editors.
From an ethics perspective, the use of explainable AI in off-target prediction is valuable as it allows scientists to identify which sequence features or genomic contexts lead an AI model to flag a potential off-target site, thereby focusing verification efforts on the most relevant risks. If a regulatory review panel can be presented with an interpretable rationale for why a certain guide RNA is deemed low-risk (for example, because the AI highlights that any sequences in the genome similar to the target have mismatches at positions proven critical for Cas9 cleavage), this transparency can facilitate informed decision-making and oversight.
In the clinical consent process, explainability is equally important. Patients enrolling in gene editing trials should be briefed on how the therapy was designed to maximize safety—here, AI-driven design can be explained in lay terms (e.g., “computer models analyzed the genome to predict and avoid unsafe targets”). Such explanations help demystify CRISPR and AI for patients, supporting autonomous informed consent by clarifying why researchers believe the approach is safe and justified. Furthermore, long-term follow-up is ethically mandated for gene editing patients to monitor for delayed adverse effects (like cancers or germline transmission in the case of reproductive cells being edited inadvertently). Explainable models can assist in post-trial monitoring by continuously evaluating patient genomic data (if available from blood or tissue samples) for signs of clonal expansion or mutations, flagging potential problems in a way that clinicians can interpret and act upon responsibly. Ultimately, maintaining trust requires that both the technology (CRISPR) and any supporting AI are used transparently and with patient welfare as the paramount concern.

\subsection{Equity and Access to Genome Editing Therapies}
A major ethical challenge as CRISPR therapies reach the clinic is ensuring equitable access to these life-saving innovations. Advanced therapies like exa-cel or CAR T cells are resource-intensive and expected to be costly, raising concerns that they may remain inaccessible to patients in low-income regions or underfunded healthcare systems, thereby exacerbating global health inequity. As Anne Muigai argues, there is an ethical imperative to expand access to genetic therapies worldwide rather than concentrating benefits in wealthy nations\cite{Muigai2022}. Strategies proposed to promote equity include tiered pricing of CRISPR therapies (charging lower prices in low-income markets), innovative financing and subsidy mechanisms, open licensing of critical patents for humanitarian use, and capacity-building in developing countries to deploy gene therapies safely\cite{Muigai2022}. The goal is to avoid a scenario where, for example, cures for SCD – a disease that affects millions in Africa and South Asia – are only available to a few hundred patients in high-income countries.
There are precedents in the pharmaceutical world for differential pricing and patent pooling for HIV and other diseases which could serve as models. In the context of CRISPR, companies and policymakers will need to collaborate to ensure that first-generation genome editing cures (like those for hemoglobinopathies) eventually become accessible beyond elite centers. Explainable AI can support these equity efforts by lowering barriers to entry for using CRISPR. Open-source AI tools for guide design and safety evaluation can empower researchers in under-resourced settings to design their own CRISPR strategies for local disease priorities without needing extensive proprietary know-how. If these tools are interpretable, they can serve as educational resources as well, helping train the next generation of scientists globally in genome editing design principles. Democratizing the AI that underpins CRISPR experiments is one step toward democratizing the therapy itself. Moreover, as regulatory agencies in different countries evaluate CRISPR trials, having transparent AI-based evidence of safety and efficacy could expedite approvals in countries that might not have the same depth of review infrastructure, provided the tools are validated. This again ties into equitable access: a globally harmonized yet transparent set of AI-assisted criteria for what constitutes a “safe enough” edit could help more countries comfortably adopt CRISPR therapy when it becomes available.

\subsection{Intellectual Property, Regulation, and Ethical Research Conduct}
The whirlwind race to patent CRISPR technologies (and the infamous patent disputes between major institutions) underscores another facet of the ethical landscape: how intellectual property (IP) rights intersect with responsible science. Patents can incentivize innovation but, if too restrictive, might stifle research or hinder access. Different jurisdictions have taken varying stances on patenting gene editing. Notably, China—an emerging leader in CRISPR research—has grappled with the moral implications of CRISPR patents. Chinese patent law includes provisions that inventions violating moral standards are not patentable, but historically it was unclear how this applied to CRISPR, particularly after the gene-edited babies incident. An analysis by Zhang (2021) discusses how Chinese authorities responded by updating patent examination guidelines to incorporate ethical reviews for CRISPR-related patents\cite{Zhang2021Patents}. This means that an application to patent a CRISPR method (for example, one involving human embryos) could be rejected on ethical grounds even if it is scientifically novel and non-obvious. The intention is to ensure that the patent system does not implicitly endorse or encourage morally contentious experiments by granting them intellectual property protection\cite{Zhang2021Patents}.
This move is part of a broader recognition that legal frameworks—patent law, regulatory approvals, research funding policies—must align with ethical norms to guide the development of CRISPR. From a regulatory standpoint, agencies are now developing specific guidelines for genome editing clinical trials. These often include requirements for additional layers of review, such as institutional biosafety committees, embryology oversight committees for any germline-related research, and data monitoring boards with genomic expertise for trials. Many countries have set up national commissions or task forces to periodically re-evaluate gene editing policies as the science evolves.
In terms of research conduct, journals and funding bodies now commonly require ethics statements and evidence of regulatory clearance for CRISPR experiments, especially those involving human cells or embryos. The role of AI in this regulatory and IP context is still emerging, but one can foresee that as AI becomes integral in designing CRISPR experiments, questions will arise about the ownership and accountability of AI-generated designs. If an AI tool recommends an edit that has unintended consequences, where does liability lie—with the tool’s developers or the scientists using it? Ensuring the AI is explainable contributes to responsible conduct by allowing researchers to double-check and rationalize AI suggestions rather than blindly trust a “black box.” It also raises the possibility of sharing the rationale for certain AI-informed design choices in patent applications or regulatory submissions, which could become part of the record that an examiner or reviewer considers. Overall, embedding ethical thinking into every stage of CRISPR innovation—from algorithm design to patient delivery—will be crucial to realizing the technology’s benefits without betraying public trust.

\section{Multi-Omics Integration and Emerging Genome Editing Techniques}
As CRISPR technology matures, it is branching out in two synergistic directions: (1) integration with other “omics” modalities and high-throughput techniques to achieve a deeper, systems-level understanding of gene function, and (2) the development of novel genome editing tools and variants (such as base editors, prime editors, and RNA-guided integrases) that expand the scope of possible genetic alterations. Both trajectories are being accelerated by contributions from AI and are yielding innovative approaches to enhance precision and efficacy. In this section, we explore how combining CRISPR with multi-omics readouts is transforming functional genomics, and we highlight emerging gene editing techniques and their implications for safety and clinical translation.

\subsection{High-Throughput CRISPR Screens and Multi-omics Analyses}
CRISPR-Cas systems have been widely adopted for genome-wide screening: by designing libraries of sgRNAs targeting thousands of genes, researchers can perturb many genes in parallel and measure phenotypic outcomes to identify genes involved in a given biological process. Initially, readouts of such screens were often crude (cell survival, fluorescence of a reporter, etc.), but a new generation of studies is coupling CRISPR perturbations with rich multi-omics single-cell readouts to glean more information from each perturbation.
One pioneering method, Perturb-seq, allows pooled CRISPR knock-out or CRISPR interference (CRISPRi) experiments to be analyzed via single-cell RNA sequencing. Replogle \textit{et al.}\ introduced a technique to capture the identity of the sgRNA present in each cell alongside that cell’s transcriptome, so that the gene expression changes caused by knocking out combinations of genes can be measured at single-cell resolution\cite{Replogle2020}. Using this approach, they interrogated combinations of epigenetic regulators in neuronal cells and discovered synergistic interactions that would be invisible to one-gene-at-a-time screens\cite{Replogle2020}. The integration of transcriptomic profiling (an “omics” layer) with CRISPR perturbations greatly expands the data output of functional screens, enabling researchers to map genetic networks and pathways with unprecedented detail. Similarly, other studies have paired CRISPR screens with proteomic readouts or metabolic phenotyping, though these are technically more complex. In all cases, the data volumes are enormous, and AI (particularly deep learning) is instrumental in pattern recognition—distilling which gene perturbations drive specific cell-state changes out of noisy single-cell data. Interpretability in these models is key to extracting biological insight; approaches such as clustering gene expression programs or using attention-based models to link sgRNA identities to transcriptional outcomes can help identify why a particular gene knock-out produces a given phenotype.
Another frontier is image-based CRISPR screening, sometimes dubbed “CRISPR microscopy.” Rather than using sequencing as the readout, these screens rely on automated imaging to visualize phenotypic changes in cells for each gene perturbation. Kanfer \textit{et al.}\ demonstrated an image-based whole-genome CRISPR interference screen to find genes regulating subcellular organelles\cite{Kanfer2021}. They targeted every gene in the genome with CRISPRi and used high-content fluorescence microscopy to quantify changes in peroxisome abundance in thousands of cells, identifying novel factors involved in peroxisome biogenesis\cite{Kanfer2021}. This method generates terabytes of image data; here, advanced image analysis algorithms (often machine learning-driven) are critical to extract features and detect subtle morphological changes. The involvement of AI in such pipelines is inevitable, and ensuring these algorithms are explainable is important for biological interpretation—researchers need to trust that the features driving a clustering of “hits” in an image-based screen correspond to real, interpretable cellular phenotypes (like organelle size or count, not an artifact of staining). By bridging microscopy (a form of phenomics) with genomics, image-based CRISPR screens exemplify multi-omic integration. They also illustrate how interdisciplinary CRISPR research has become, merging genome editing, cell biology, and AI-based image informatics.
Advancements have also been made in tuning the magnitude of CRISPR perturbations to study dosage-dependent effects. Traditional CRISPR knockouts abolish gene function completely, but biology is often not binary. Jost \textit{et al.}\ developed libraries of “titrated” sgRNAs that produce gradations of gene repression or activation by CRISPRi/a, achieved by introducing mismatches into the sgRNA to systematically reduce its binding efficiency\cite{Jost2020}. This allowed them to generate fine dose-response curves of gene function within a single screen, essentially merging quantitative genetics with CRISPR screening\cite{Jost2020}. Multi-omics readouts, such as measuring both transcriptomic and phenotypic changes at different knockdown strengths, can further enrich these datasets. AI models that can integrate these multiple data types (perturbation dose, gene expression changes, cell images, etc.) will be crucial for fully exploiting such rich experiments. An explainable multi-modal model could potentially reveal, for example, that partial inhibition of a certain essential gene triggers a compensatory transcriptional program, whereas full knockout triggers cell death, pinpointing thresholds that might be therapeutic sweet spots.
Through these examples, it is clear that combining CRISPR with multi-omics is yielding deeper insight into gene function and interactions. The data complexity demands computational assistance, and indeed many of these studies are as much a triumph of bioinformatics and AI as of genome editing. Importantly, the synergy goes both ways: while AI helps make sense of CRISPR screens, the data from these screens in turn feed back into AI models, improving their knowledge of gene networks and the consequences of perturbations. This virtuous cycle will enhance our ability to predict the outcomes of gene edits in different cellular contexts, a necessary step for safely moving genome editing into more complex therapeutic areas.

\subsection{Emerging Gene Editing Technologies: Base Editing, Prime Editing, and Beyond}
In parallel to applying CRISPR in innovative ways, researchers have been inventing new classes of genome editors to address limitations of the classic CRISPR-Cas9 nuclease. Base editors and prime editors are two such technologies that have rapidly advanced in the past few years, enabling precise nucleotide changes without creating double-strand DNA breaks. Base editors (BEs) are fusion proteins that couple a DNA-modifying enzyme (like cytidine deaminase or adenosine deaminase) to a catalytically impaired Cas enzyme, allowing specific base conversions (C$\rightarrow$T or A$\rightarrow$G, respectively) at targeted sites guided by a sgRNA. Prime editors (PEs) extend this concept by fusing a reverse transcriptase to Cas9 and using a complex guide RNA (the pegRNA) that both targets the site and encodes a desired edit in an RNA template, which gets copied into the genome\cite{An2024PrimeEVLP}. These technologies can potentially correct point mutations or introduce small insertions/deletions in a programmable way, with fewer byproducts than nuclease-induced random indels.
A flurry of research has improved the efficiency and delivery of base and prime editors. For base editing, a key hurdle was packaging the editor into viral vectors for in vivo use, given the size of Cas9 plus the deaminase. Davis \textit{et al.}\ engineered a hypercompact adenine base editor and optimized the genome of adeno-associated virus (AAV) such that the entire editor could fit in one AAV particle\cite{Davis2022ABE}. This enabled efficient delivery of an AAV encoding a base editor to mouse liver, where it introduced a precise A$\to$G mutation in the \textit{PCSK9} gene, resulting in lowered cholesterol levels in the animals\cite{Davis2022ABE}. The one-vector base editor system achieved high editing rates and had minimal off-target DNA or RNA edits, demonstrating the progress in making base editing clinically viable.
For prime editing, initial versions were limited by their cargo size and the challenge of delivering an RNP that includes a long pegRNA. In 2024, two complementary advances were reported: Davis \textit{et al.}\ showed that “split” prime editors delivered via dual AAVs could mediate efficient editing in multiple mouse tissues (including brain, liver, and heart) with editing frequencies up to 40–50\% and no detectable off-target events\cite{Davis2024Prime}; and An \textit{et al.}\ developed engineered virus-like particles (eVLPs) that package the prime editor protein and pegRNA as a ribonucleoprotein, enabling transient delivery of prime editing machinery to cells without any DNA vector\cite{An2024PrimeEVLP}. The eVLP approach successfully corrected a pathogenic mutation in a mouse model of inherited blindness, achieving therapeutically relevant levels of editing with minimal off-target activity and avoiding the risk of vector integration since no DNA is introduced\cite{An2024PrimeEVLP}. These innovations significantly broaden the disease contexts in which prime editing could be applied—high efficiency in vivo editing means one could conceivably correct mutations in post-mitotic tissues like muscle or brain, which is highly relevant to diseases such as muscular dystrophy or neurodegenerative disorders.
Beyond base and prime editing, entirely new CRISPR-based platforms are emerging. One such platform is CRISPR-mediated integrases for large DNA insertions. Traditional CRISPR is efficient at disrupting genes or making small fixes, but inserting a new gene (say, to replace a defective one) has been challenging because it relies on cellular repair mechanisms like homologous recombination. Yarnall \textit{et al.}\ introduced a method called PASTE (Programmable Addition via Site-specific Targeting Elements) which couples prime editing with site-specific serine integrases to “drag-and-drop” kilobase-sized DNA sequences into the genome without requiring double-strand breaks or homology-directed repair\cite{Yarnall2023PASTE}. In human cells, PASTE achieved integration of whole gene cassettes (on the order of 4–10 kb) at specified loci with ~20\% efficiency\cite{Yarnall2023PASTE}. This represents a significant technological leap: instead of just correcting typos in the genome, we can now consider pasting in entire new paragraphs of DNA text. The potential applications are vast, from installing therapeutic genes in patients with monogenic disorders to inserting safety switches or logic circuits into cell therapies. However, these new tools also introduce new variables—multiple components (Cas9, reverse transcriptase, integrase in the case of PASTE) and more complex target-site requirements—which again will benefit from AI to guide their use. For instance, selecting optimal attB/attP recombination sites for integrase action across a patient’s genome is a task well-suited for predictive modeling, and one must balance efficiency with the risk of integrating into a hazardous location. Explainable models could help flag, for example, if a chosen safe-harbor locus for integration has any unforeseen regulatory importance.
Crucial to all these emerging editors is the question of off-target effects and safety. Interestingly, because base editors and prime editors do not induce a blunt DNA cut, they tend to have different off-target profiles (base editors can cause off-target point mutations, including RNA edits in some cases; prime editors can have byproducts like small indels at the target site). Extensive profiling is underway: methods like high-throughput DNA sequencing and sensitive off-target detection adaptations (e.g., modifying DISCOVER-Seq or CHANGE-seq to detect base-editing events) are being applied. One study enhanced the sensitivity of the in situ Cas9 activity detection method DISCOVER-Seq by adding a pre-amplification step (creating “DISCOVER-Seq+”), which made it possible to detect rare off-target edits even in vivo, increasing the number of detectable off-target sites several-fold\cite{Wiener2023DISCOVERSeqPlus}. Diligent application of such methods to base and prime editors will be key to mapping their specificity. AI models trained on these data can then predict off-target propensities of different editors, ideally providing not just a yes/no but an interpretable reason (e.g., “this off-target site has a protospacer adjacent motif compatible with the nickase in the prime editor and a $3^\prime$ microhomology to the pegRNA extension, making it susceptible to unintended editing”). Early results are encouraging: prime editors have so far shown low off-target rates in vivo when properly delivered\cite{Davis2024Prime}, and base editors have been refined to minimize RNA off-target by engineering the deaminase component.
Yet another dimension is the exploration of novel Cas variants and other CRISPR-associated systems to complement SpCas9. Some Cas9 variants and orthologs can target different PAM sequences, which expands the range of genomic sites that can be edited. Zhang \textit{et al.}\ undertook a comprehensive assessment of several engineered Cas9 variants like xCas9 and SpCas9-NG (which recognize more relaxed PAMs than the canonical NGG) as well as other orthologs, profiling their activities on large target libraries\cite{Zhang2021Cas9Variants}. They charted which variants were most efficient for various non-NGG PAM sites and also identified idiosyncrasies in their off-target profiles\cite{Zhang2021Cas9Variants}. Such data serve as a valuable resource; for any given target sequence, an algorithm could recommend the best Cas nuclease variant that balances on-target efficacy and specificity.
In effect, the CRISPR toolbox is diversifying—Cas12a (Cpf1) brings alternate cut patterns and a different PAM, Cas13 enables RNA editing, and completely new systems like Cas-associated transposases are on the horizon. AI plays a particularly important role in mastering this growing toolbox. With so many options (different editors, delivery methods, target choices), designing an optimal therapeutic strategy becomes a high-dimensional optimization problem. Machine learning models can be trained on past experimental results to predict which strategy is likely to yield the highest editing efficiency with lowest risk for a new target. For example, given a genetic disease mutation, one might have options: use a base editor to directly fix the point mutation, or use prime editing to correct it, or perhaps knock out a modifier gene to compensate for it. Computational frameworks can evaluate these alternatives by aggregating data on what has worked for similar scenarios. If these frameworks are explainable, they can provide human designers with insights, such as “the mutation occurs in a genomic context (e.g. a repetitive sequence or a structured chromatin region) where base editing efficiency is typically low, hence prime editing is recommended despite being slower, because the model’s learned parameters indicate prime editing is less affected by that context.” Such reasoning makes the recommendation more trustworthy and scientifically grounded.

\begin{figure*}[ht] 
\centering
\caption{\textbf{Emerging CRISPR-based technologies and multi-omics integration are extending the frontiers of genome editing.} \textbf{(A)} \textit{Base editing} (left) and \textit{prime editing} (right) enable precise nucleotide changes without cutting both DNA strands. Base editors (e.g., adenine base editors) have been optimized for in vivo delivery by using smaller Cas9 variants and deaminases, allowing single-AAV delivery and successful gene editing in mouse models (for instance, lowering \textit{PCSK9} to reduce cholesterol)\cite{Davis2022ABE}. Prime editors use a pegRNA and a Cas9-RT fusion to “write” new DNA sequences; recent advances like dual-AAV split prime editors achieved high editing rates (up to 40–50\%) in multiple tissues in vivo\cite{Davis2024Prime}, and non-viral delivery via eVLPs has shown efficient correction of disease mutations with minimal off-target events\cite{An2024PrimeEVLP}. \textbf{(B)} Large-sequence insertions via CRISPR: novel techniques such as \textit{PASTE} combine prime editing with integrases to insert kilobase-sized DNA payloads at specified genomic sites\cite{Yarnall2023PASTE}. This “drag-and-drop” integration can add entire genes (e.g., to replace a defective gene or add a new function) without requiring double-strand breaks, greatly expanding potential therapeutic applications. \textbf{(C)} Multi-omics CRISPR screens: high-throughput perturbation is paired with rich data readouts, such as single-cell RNA sequencing in Perturb-seq or high-content imaging. This allows researchers to observe not just whether a cell lives or dies after a gene knockout, but exactly how gene networks and phenotypes shift in response\cite{Replogle2020,Kanfer2021}. AI methods integrate these complex datasets to map genetic interactions; importantly, explainable models help identify causal links (for example, highlighting which transcriptional changes are directly attributable to a given gene edit). \textbf{(D)} Enhanced precision and targeting: a growing arsenal of Cas enzyme variants (each recognizing different PAM sequences or having altered cutting properties) and auxiliary methods for off-target detection are improving the precision of genome editing. For instance, engineering of SpCas9 variants like xCas9 and Cas9-NG has expanded targetable sites beyond the NGG PAM constraint\cite{Zhang2021Cas9Variants}, while advanced off-target detection techniques (e.g., DISCOVER-Seq+ in vivo\cite{Wiener2023DISCOVERSeqPlus}) ensure that even rare off-target events can be detected and mitigated. These developments, combined with AI-driven design algorithms, enable a more tailored and safe application of genome editors for any given genomic locus.}
\label{fig:emerging}
\end{figure*}

\section{Conclusion}
The synergy between next-generation gene editing and explainable AI is poised to fundamentally reshape the landscape of biotechnology and medicine. As CRISPR-based therapies transition from experimental endeavors to approved clinical treatments, the demands on precision, safety, and predictability have never been higher. AI systems, especially those endowed with explainability, have emerged as essential partners in meeting these demands—guiding the design of sgRNAs and novel editors, predicting and interpreting off-target effects, and distilling complex multi-omics data into actionable knowledge. This partnership is inherently interdisciplinary: it requires deep knowledge of molecular biology, computational prowess, and ethical insight in equal measure.
By integrating explainable AI at each step of the genome editing pipeline, researchers can iterate faster and more safely. In silico models can rapidly screen countless potential edits and flag the most promising ones, reducing costly trial-and-error in wet labs. The explainability of these models ensures that their recommendations come with intelligible justifications, which is key for the end-users—the scientists and clinicians—to trust and refine the predictions. For example, rather than just outputting “Guide RNA X is optimal,” an explainable tool would indicate, “Guide RNA X is optimal because it targets an open chromatin region and avoids an off-target with a single nucleotide mismatch at a critical position,” aligning with human expertise and allowing human oversight. This transparency not only aids researchers but is also becoming crucial for regulatory approval: agencies are more likely to accept AI-informed arguments if they can understand the basis of those arguments.
The case studies discussed—from curing blood disorders and blindness to innovating new editors and safeguards—collectively illustrate a trajectory: we are moving toward a future where gene editing is routine and reliably safe. But we are also reminded of the care needed in this journey. Unintended consequences, whether molecular (like unexpected mutations) or societal (like unequal access or ethical breaches), remain real risks. Explainable AI contributes to mitigating the former by illuminating the dark corners of genome editing outcomes, and can even aid in addressing the latter by informing policymakers with clear data and predictions. Still, technology alone cannot resolve ethical dilemmas; a global commitment to responsible innovation, public engagement, and equitable access must accompany the technical advances.
Reproducibility and validation will remain cornerstones. AI predictions and novel editing methods must be rigorously validated experimentally, and results (whether successes or failures) should be reported openly to refine the collective understanding. Initiatives to create open databases of CRISPR off-target effects, or global registries of gene editing trials (as recommended by the WHO\cite{WHO2021}), are invaluable for feeding data back into AI models, making them more robust and generalizable. As models become more accurate and explainable, they will further accelerate the development of safer editors, creating a positive feedback loop.
The role of explainable AI in enhancing CRISPR precision and safety is that of an empowering enabler—amplifying human ability to engineer biology with foresight and care. Together, advances in AI and genome editing are transforming what was once science fiction into a new standard of care for genetic diseases. The coming years will likely witness the first widely available cures stemming from CRISPR technology, and these will owe much to the unseen algorithms that helped chart the course. By embracing explainability, we ensure that this voyage of discovery remains transparent and guided by understanding at every step. The result is not only more trustworthy science but also technology that society can embrace with confidence. As we stand at this nexus of AI and gene editing, the path forward is one of immense promise: a future where we can correct genetic errors as routinely as administering a drug, informed by AI-driven wisdom and governed by a steadfast commitment to ethical responsibility.
\section*{Acknowledgments}
The authors used generative AI tools for editing and formatting, but take full responsibility for the final content.

\providecommand{\bibinfo}[2]{#2}
\providecommand{\urlprefix}{URL }
\providecommand{\doi}[1]{DOI~\discretionary{}{}{}#1}

\end{document}